\documentclass[twocolumn,showpacs,preprintnumbers,amsmath,amssymb,aps]{revtex4}
\usepackage{graphicx}

\begin{document}

\title{Coarsening of Topological Defects 
in Oscillating Systems with Quenched Disorder}   
\author{C. Reichhardt and C.J. Olson Reichhardt} 
\affiliation{ 
Theoretical Division and Center for Nonlinear Studies,
Los Alamos National Laboratory, Los Alamos, New Mexico 87545}

\date{\today}

\begin{abstract}
We use large scale simulations to study interacting particles 
in two dimensions in the
presence of both an ac drive and quenched disorder.
As a function of ac amplitude, there is 
a crossover from a low drive regime where the colloid positions are 
highly disordered to a higher ac drive regime where the system dynamically
reorders.     
We examine the coarsening of topological
defects formed when the system is quenched from 
a disordered low ac amplitude state to
a high ac amplitude state. 
When the quench is performed close to the disorder-order crossover, the defect
density decays with time as a power law with $\alpha = 1/4$ to $1/3$.   
For deep quenches, in which the ac drive is increased to high values
such that the dynamical shaking temperature is strongly reduced,
we observe a logarithmic decay of the defect density into a grain boundary 
dominated state.  We find a similar logarithmic decay of defect density
in systems containing no pinning.
We specifically demonstrate these effects for
vortices in thin film superconductors, and discuss implications for
dynamical reordering transition studies in these systems.    
\end{abstract}

\maketitle
\vskip2pc

\section{Introduction}

There are a wide variety of systems that can be modeled 
as an elastic lattice driven over an underlying random quenched
substrate by an external force.
Examples include moving vortex lattices in type-II superconductors
\cite{Koshelev1,Zimanyi2,Marley3,Pardo}, 
sliding charge density wave systems \cite{Balents4}, 
models of atomic friction \cite{Braun5}, 
colloidal particles moving over a rough substrate \cite{Colloid6}, 
driven Wigner crystals, and stripe forming systems \cite{Reichhardt7}.  
Typically, when the quenched disorder is strong, 
the elasticity of the system breaks down and 
topological defects proliferate in 
the form of dislocations and disclinations.     
It has also been shown 
that when a sufficiently large external drive
is applied, the system can exhibit
a dynamical reordering process where the 
density of topological defects is sharply reduced in the
presence of a high dc drive 
\cite{Koshelev1,Zimanyi2,Pardo}. 
Conversely, if the external drive is reduced
from a high value where the system is 
ordered, the underlying disorder 
acts as a fluctuating force on the moving particles
which increases as the 
drive is lowered, causing the system to dynamically melt via
the proliferation of defects \cite{Koshelev1}. 
This dynamical reordering effect has
been demonstrated for vortices in superconductors through
simulations \cite{Koshelev1,Zimanyi2},  
transport \cite{Marley3}, and imaging experiments \cite{Pardo}, 
and has also been verified for
sliding charge density waves \cite{Balents4}, colloids \cite{Colloid6}, and
electron crystals \cite{Reichhardt7}. 

Experiments have produced evidence that an applied ac drive can 
also induce a dynamical ordering transition in vortex 
systems \cite{Bekeris}. 
Further, simulations
with various types of ac drives have shown that the vortex
lattice can reorder and that the reordered 
lattice is aligned in the direction of the
ac drive \cite{Valenzuela}.  

Since these driven systems show both and ordered and disordered phases
as a function of external drive, it is also possible to study the
dynamics of a quench from either the ordered regime to the disordered
regime or the converse. 
An open question is how the density of topological defects 
decays if the system is quenched from a
regime where the steady state is disordered to
a regime where it is ordered, such as by suddenly 
changing an external ac or dc drive.  

The decay of 
topological defects after a quench has been studied 
in 
systems without quenched disorder
through
both simulations 
\cite{Elder8Cross,Mazenko9,Vinals10,Boyers11} 
and experiments 
on convection and electroconvection in 
fluids \cite{Dennin12,Carina13}.
In general in these systems, the quench is performed between ordered 
and disordered phases that are separated by a continuous phase transition.  
A number of these studies find 
that the density of defects, or of
grain boundaries which are comprised 
of defects, decays as a power law $t^{-\alpha}$, where 
the exponent $\alpha = 1/5$, $1/4$ \cite{Elder8Cross,Dennin12}, 
or $1/3$ \cite{Mazenko9,Vinals10,Boyers11,Carina13}. 
The different exponents that are obtained
depend on the model of the depth of the quench.

It has also been shown that in some cases the defects 
form  grain boundaries, and that there is a tendency for
the system to create its own intrinsic pinning of the grain boundaries 
\cite{Vinals10}.   
This can lead to glassy effects \cite{Vinals10}, 
and in this case the defect density
decays very slowly or logarithmically with time rather than 
as a power law.    
Recent experiments with colloidal particles which form
a triangular lattice in equilibrium showed that in a system with a
large number of defects, the dislocations coarsen into 
grain boundaries and the defect density  
decays logarithmically with time rather than as a power law \cite{Wu14}.
Simulations of two-dimensional vortex lattices driven with ac drives over
quenched disorder also provided evidence that the defect density 
decays logarithmically with time; however, this data was very 
noisy \cite{Valenzuela}. 

We note that in the absence of a dc or ac drive, it has been found that
a vortex lattice does not reorder into a crystal in the 
presence of disorder.  Instead, if the disorder is weak, 
the system is dominated by grain boundaries whose motion
becomes frozen \cite{Chandra,Zapperi}.  Similar behavior occurs
for a small number of 
strong pinning sites \cite{Dasgupta}.    

Here, we address how the density of topological defects 
decays  
in the presence of quenched disorder when the ac driving force is suddenly 
changed. Such a system can be
realized readily in experiment
for vortices interacting with random disorder and
an oscillatory applied current. 

Simulations of quench dynamics generally 
employ continuum models such as the Swift-Hohenberg model. 
A complementary approach 
which we adopt here is to model the individual particles of the elastic
lattice directly, and allow the topological defects to emerge naturally
from particle rearrangements.
The drawback 
to such an approach 
is that very large systems 
are needed to produce reasonable
statistics for the defect densities. 
Recent advances in computational speed have made it possible to 
access the defect density over long periods of time with particle based 
simulations and obtain defect numbers that are comparable with those of 
other simulation methods.  An additional benefit of the particle based 
method is that processes such as effective pinning due to the
ordering of the system into a triangular lattice are readily captured.   

\section{Simulation}

We consider elastic lattices interacting with
quenched disorder in two dimensions for the specific cases of
vortices in thin-film type II superconductors and charged colloids. 
In both cases, the equilibrium configuration of the $N=6000$ 
repulsive overdamped particles
in the absence of disorder and 
temperature 
is a defect-free triangular lattice.
Particle $i$ obeys the overdamped equation of  
motion  
\begin{equation}
\frac{d {\bf r}_{i}}{dt} = \sum^{N}_{j \neq i}{\bf f}_{ij}^{pp} + 
{\bf f}_{p} + {\bf f}_{AC} 
\end{equation}
Here the interaction force 
between particles a distance $r_{ij}$ apart is 
${\bf f}_{ij}^{pp} = -\nabla_i U(r_{ij})\hat{\bf{r}}_{ij}$,
where $U(r_{ij}) = \ln(r_{ij})$ for vortices in superconducting films, 
and $U(r_{ij}) = \exp(-r_{ij})/r_{ij}$  for colloids. 
The force ${\bf f}_{p}$ from the disordered substrate is 
exerted by $N_{p}$ randomly placed parabolic pinning sites of
radius $r_{p}=0.2$ and strength $f_{p}$. 
An external 
ac force ${\bf f}_{AC}=A\sin(\omega t)\hat{\bf x}$ 
is applied to all particles.
We consider only the 
$T=0$ case, 
so that the only source of intrinsic noise is the 
fluctuations 
caused 
when the particles move over the quenched 
disorder.

\begin{figure}
\includegraphics[width=3.5in]{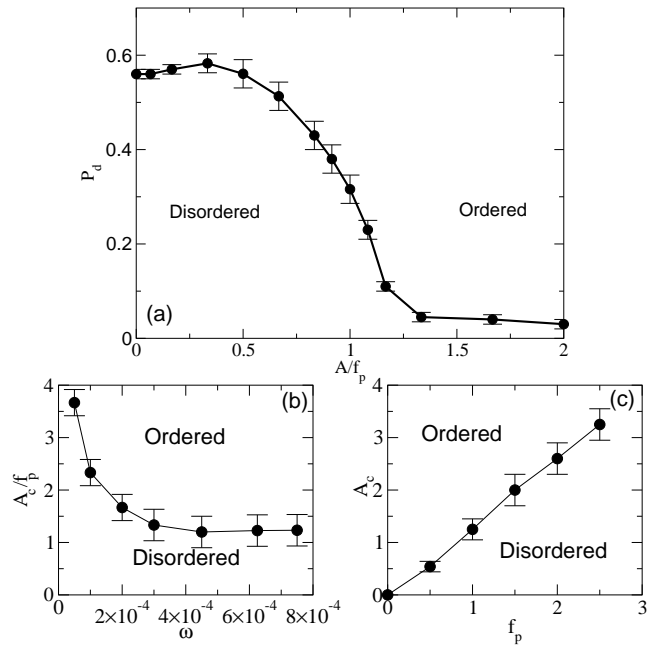}
\caption{(a) $P_{d}$ vs 
$A/f_{p}$, for fixed $f_{p}$=1.5 and $\omega=3\times 10^{-4}$.    
(b)  $A_c/f_{p}$ 
vs $\omega$ at fixed $f_{p}=1.5$. (b) $A_c$ vs $f_{p}$ 
at fixed $\omega=3\times 10^{-4}$. 
}
\end{figure}

\section{Dynamical Reordering with AC drives} 

We first identify where the disordered and ordered phases occur as a function
of pinning strength, pinning density,
and ac drive parameters. 
We slowly increase the magnitude of the ac force, $A$, and
measure the density of topological defects by determining the fraction 
$P_d$ of particles that are not sixfold coordinated.
In a triangular lattice, $P_{d} = 0$.
The regimes we observe 
can also be distinguished via time-dependent 
fluctuations in $P_d$.
At low $A$, a disordered 
pinned phase appears in which the particles remain
mostly immobile and $P_d$ is roughly constant with time. 
For intermediate $A$, there is a disordered plastic flow phase 
where only a portion of the particles are moving.
At high $A$, the system orders into a smectic type state where the
lattice aligns in the direction of the drive and there are only 
a small number of dislocations present which 
have aligned Burgers vectors. 
These results are consistent with those obtained in simulations with slowly 
increasing or decreasing dc drives \cite{Koshelev1,Zimanyi2}. 

In Fig.~1(a) we plot the density of defects $P_{d}$ vs $A/f_p$ 
for superconducting vortices 
with fixed pinning density $\rho_p=1.3$, 
pinning strength $f_p=1.5$, and ac drive frequency $\omega=3\times 10^{-4}$. 
We normalize $A$ by $f_{p}$ so that the crossover
from the disordered phase to the ordered phase 
occurs near $A/f_{p} = 1.0$. 
For 
$ A/f_{p} < 1.0$ the system is 
strongly disordered 
and a large fraction of the particles are not sixfold coordinated. 
For $A/f_{p} > 1.3$ the system orders and the density of defects is 
strongly reduced. The disorder to order crossover occurs for a
wide range of 
parameters. In Fig.~1(b) we plot the crossover line 
$A_c/f_p$ 
from the
disordered to the ordered state vs ac drive frequency $\omega$
for fixed $f_{p}$.
As $\omega$ is lowered, a larger $A$ must be applied to order the system. 
For high $\omega$, the crossover saturates near $A_c/f_p = 1.3$. 
In Fig.~1(c) we plot $A_c$ vs $f_{p}$ at fixed $\omega$. 
Here, $A_c$ increases roughly linearly with $f_{p}$.

\begin{figure}
\includegraphics[width=3.5in]{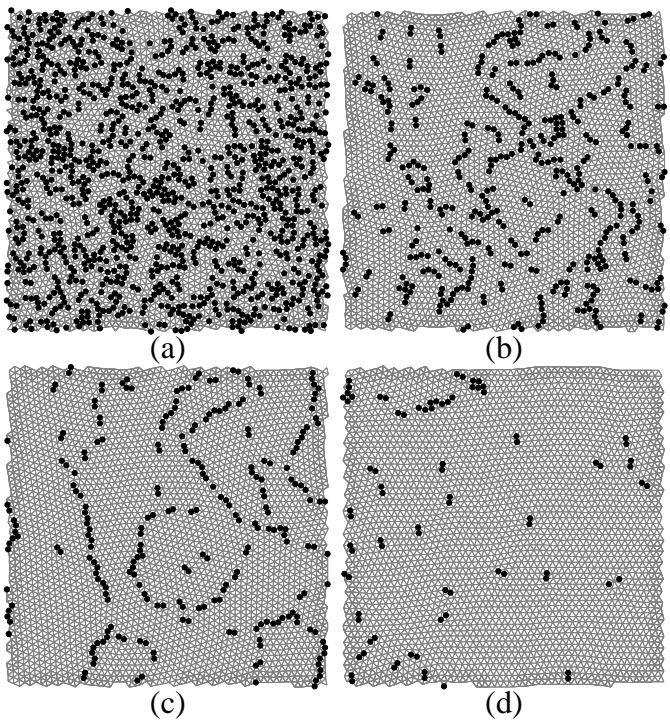}
\caption{
Delaunay triangulation showing defect coarsening
for the system in Fig.~1(a) at $A^*/f_p=2.0$. 
(a) Initial state at $t=0$; (b) $t = 1\times 10^4$; (c) 
$t = 7\times 10^4$;  (d) $t = 4.5\times 10^5$. 
}
\end{figure}

\section{Defect Decay After a Quench of the AC Drive} 

We study
the effect of a sudden quench by 
abruptly changing $A$ from the 
low drive disordered regime 
to a high drive value $A^*$ where the steady state would be ordered.      
The defects present in the disordered phase coarsen after $A$ is changed.
We first consider the case $A^* \gtrsim A_c$.
In Fig.~2 we show the 
real space time 
evolution of the coarsening of the defects for the system in 
Fig.~1(a) with $f_p=1.5$ where $A/f_{p} = 0.25$ 
is suddenly changed to
$A^*/f_p=2.0$. In this case,
$A_c/f_p = 1.3$.  
The initial strongly disordered state is illustrated in Fig.~2(a), and
the defects in the Delaunay triangulation are 
highlighted in black.  
In Fig.~2(b) at time $t = 1\times 10^4$ 
a large portion of the defects have annihilated 
and some grain boundary structures appear. 
In Fig.~2(c) at $t = 7\times 10^4$ the ordered
domains are larger, while in Fig.~2(d)
($t = 4.5\times 10^5$) 
all 
of the grain boundaries are gone and 
only a small number of isolated dislocations are present. 
The system is also aligned in 
the $x$ direction with the ac drive.
The remaining isolated dislocations do not annihilate with time when we
continue the simulation over a longer interval.  

\begin{figure}
\includegraphics[width=3.5in]{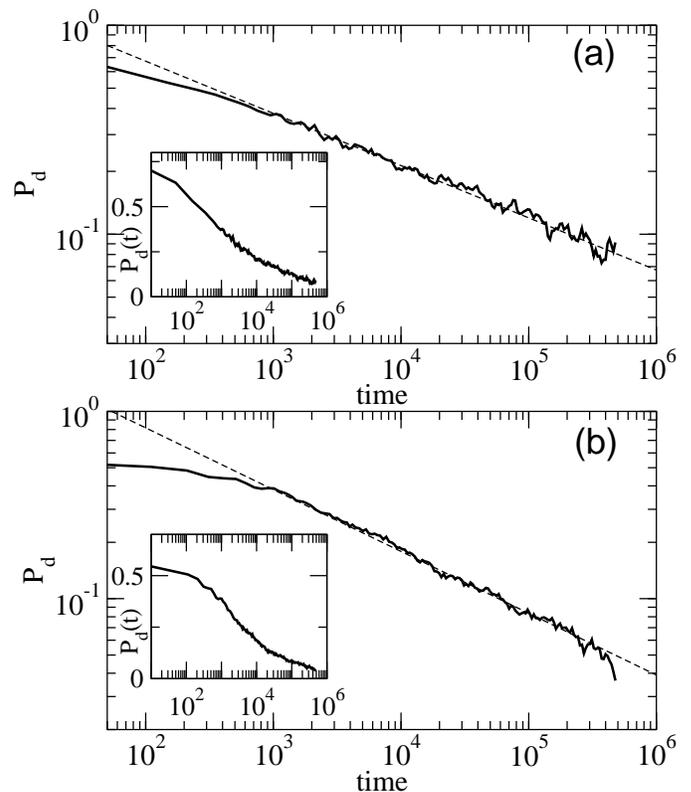}
\caption{(a) 
$P_{d}(t)$ for $A^*/f_p=2.0$ for the system in Fig.~2 with $f_p=1.5$.
Dashed line: power law fit to $1/t^{\alpha}$ with $\alpha = 0.25$.   
Inset: linear-log plot. 
(b) $P_{d}(t)$ for the same system with $A^*/f_p = 4.0$.
Dashed line: power law fit with $\alpha = 0.33$. 
Inset: linear-log plot.    
}
\end{figure}

In Fig.~3(a) we plot the time dependent defect density 
$P_{d}(t)$ 
for the system in Fig.~1(a) and Fig.~2. Here $P_{d}(t)$ 
decays as 
$1/t^\alpha$ at long times with $\alpha$ very close to $1/4$. The dashed line
is a fit over three decades with $\alpha = 0.25$.
Since the exponent is small, it may be difficult to distinguish 
a logarithmic decay from a power law decay. 
In the inset to Fig.~3(a) we plot $P_d(t)$ on a linear-log scale which would
give a straight line if the decay was purely logarithmic. 
A logarithmic decay can only be fit at shorter times, whereas the power law
fit is clearly better at the longer times.
In Fig.~3(b) we show $P_{d}(t)$ for the same system 
in Fig.~1(a) when $A^*/f_p = 4.0$. In this case we find a similar long time
power law decay of $P_{d}(t)$; however, the decay is faster 
than in Fig.~3(a) with $\alpha = 0.33$ as
indicated by the dashed line. In the inset 
of Fig.~3(b) we plot the 
curve on a linear-log scale, which 
shows that the long time behavior is consistent with a power law rather than a
logarithmic 
decay. 
As $A^*$ is increased further, the 
exponent $\alpha$ remains 
close to 0.33; however,
at very large $A^*$ the grain boundaries do not completely anneal. 

\begin{figure}
\includegraphics[width=3.5in]{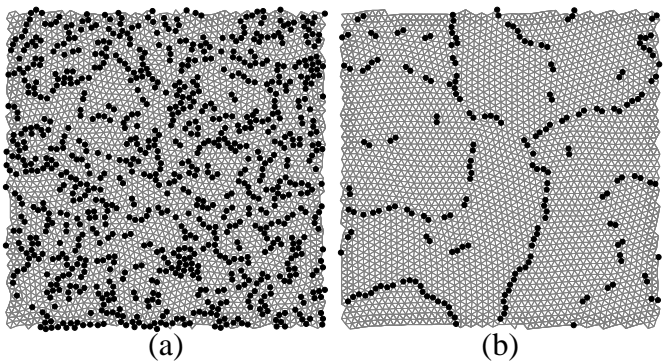}
\caption{
Topological defects (black dots) for the system in Fig.~1(a) with 
$A^*/f_p = 12$. (a) $t = 200$; 
(b) $t = 2\times 10^5$. 
}
\end{figure}

The power law decay in the defect density with
exponent $ 1/5 < \alpha < 1/3$ is  consistent with 
the exponents observed in quench studies of
continuum simulations \cite{Elder8Cross,Mazenko9,Vinals10,Boyers11}  
and convection experiments \cite{Dennin12,Carina13}. 
The change of the exponent with $A^*$ likely 
reflects the different effective 
pinning 
and defect production rates for different values of 
$A^*$. 
Studies of particles 
driven with a dc force over quenched disorder 
have found that even at $T = 0$ the particles 
experience an effective shaking temperature $T_{s}$ 
due to the randomness of the disorder. 
This shaking temperature varies inversely with 
the average particle velocity, $T_{s} \propto 1/v$ \cite{Koshelev1}, so at 
higher velocities the shaking effect is reduced. 
When $A^*$ is only slightly higher than $A_c$,
the effective
temperature is high enough both to create and to destroy defects, so the
overall rate of defect annihilation is slower. 
For higher $A^*$ values, the shaking temperature is lower,
which reduces the rate at which new defects are created; however, the
shaking temperature is still large enough to permit a significant amount
of thermally activated defect hopping to occur, 
so the net rate of defect annihilation is higher.
We note that if $A^*$ is very large, 
the net defect annihilation rate decreases
because the defect mobility from the shaking temperature is
considerably reduced.

When $A^*$ is much larger than the 
crossover value, the effective 
shaking temperature is very low since
the average velocity of the particle is larger. 
Additionally,
it was shown in the work of 
Boyer and Vi\~{n}als \cite{Vinals10} that defects and grain boundaries can
be pinned effectively by the underlying pattern that the system forms. 
This pinning can lead to long time
configurations that do not completely 
order and have very slow dynamics.
For the particular system
studied in Ref.~\cite{Vinals10}, the pinning was caused when the
ordered regime formed stripes. In our
system the
ordered phase is a triangular lattice rather than a stripe pattern;
however, this 
still generates an 
effective periodic potential through which
the defects and 
grain boundaries move. 
When the effective shaking temperature is low,
activated motion of defects is suppressed. 
In Fig.~4 we show the real space evolution of the coarsening of the defects 
for the same system in Fig.~1(a) with 
$A/f_p = 0.25$ quenched to
$A^*/f_p = 12.0$. 
In Fig.~4(a) we illustrate the configuration close to the start of 
the quench at $t = 200$ where a large number of defects are present. 
At long times, the defect density saturates and a grain 
boundary network 
remains, as shown in Fig.~4(b) 
for $t = 2\times 10^5$. 
These grain boundaries show no motion over 
extended periods of time, indicating that they
are effectively pinned by the potential created by the triangular 
ordering of the particles.  

\begin{figure}
\includegraphics[width=3.5in]{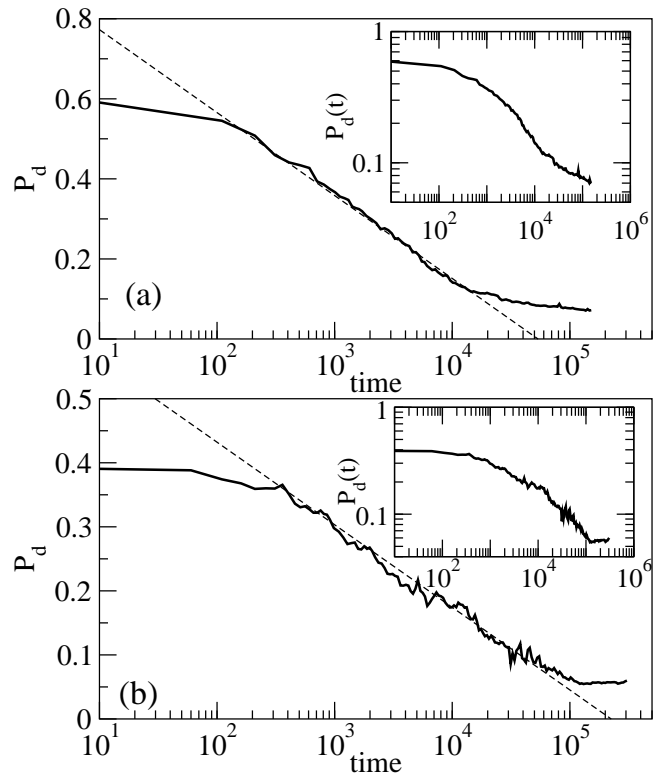}
\caption{
$P_d(t)$ for the same system
as in Fig.~1(a) at $A^*/f_p=12$.
Dashed line: logarithmic fit. Inset: log-log plot. 
(b) $P_{d}(t)$ for a system  with 
the same parameters as in Fig.~5(a) with colloidal particles.
Dashed line: logarithmic fit. Inset: log-log plot. 
}
\end{figure}

In Fig.~5(a) we plot $P_{d}(t)$ for the same system in Fig.~1(a) and
Fig.~4,
where $A^*/f_{p} = 12.0$. Here the decay is not well fit by a
power law, as shown by the inset.
There is a regime at intermediate times 
that can be fit to a logarithmic decay with $P_{d}(t) =  a\ln(t) + b$,
as indicated by the dashed line.     
At longer times, the decay slows as the grain boundaries form.
Quenches with even higher $A^*$ give very similar results.  
We find similar behavior for the case of colloidal particles, as illustrated
in Fig.~5(b) for the same parameters;
however, we observe a much stronger arrest of the decay at 
long times. 
The final configurations in the colloidal case
are also grain boundary structures. In the inset of Fig.~5(b) a
log-log plot 
indicates that the data cannot be fit by a power law.
Recent experiments with colloidal particles in two dimensions where a
mechanical shaking was applied produced a 
logarithmic decay in the length of the grain boundaries over  two decades
of time.
In some cases, a long time defect saturation has also been
observed \cite{Wu14}.     
We note that for smaller $A^*$ 
in the colloidal system, we obtain the same power law 
decays found for the vortex system. 

\section{Effects of Varied Disorder Strength, ac Drive Frequency, 
and System Size} 

We next consider the effects of varying other parameters. 
We focus on the case 
with $A/f_p = 4.0$ where there is a power law decay with
an exponent close to $0.25$. 

We first consider whether the exponent for the decay of the
defect density is affected by 
variations in the system size. In Fig.~6 we    
plot the defect decay for systems with the same parameters as
in Fig.~3(b), $f_p=1.5$, $A^*/f_p=4$, and $\omega=3\times 10^{-4}$ 
for
sizes of $L = 48$, 36, and  $24$.
The dashed line indicates a fit to $1/t^\alpha$ for $\alpha = 0.25$. 
The exponent does not change with system size, although 
the smaller system shows more fluctuations due to the 
smaller overall number of defects. 
This indicates that the system sizes we are using are large enough
to avoid finite size effects.

\begin{figure}
\includegraphics[width=3.5in]{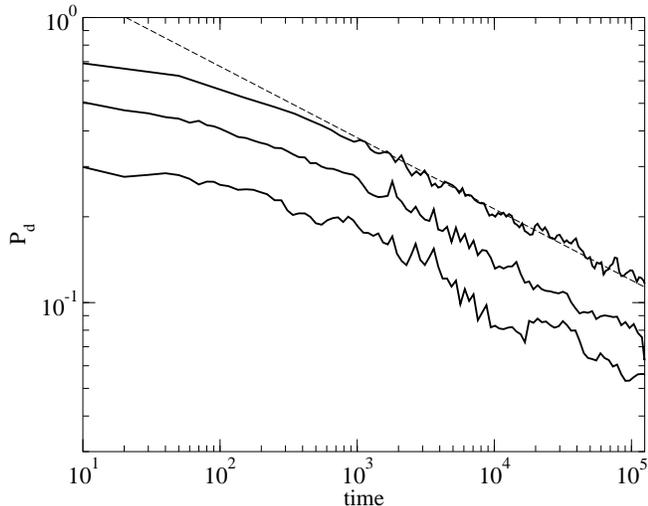}
\caption{
$P_d(t)$ for the same system
as in Fig.~3(a) for varied system sizes $L = 48$ (top), 36
(center), and $24$ (bottom).
The straight line is a 
power law fit to $1/t^{\alpha}$ with $\alpha = 0.25$.  
}
\end{figure}

We next consider the effect of changing the disorder strength $f_p$. 
As noted earlier, if the final ac amplitude is large, 
the defect decay appears to have a logarithmic form and
the system reorders to a grain boundary dominated state 
rather than to a state where all the
remaining defects are aligned in the direction of the ac drive. 
This occurs when the final ac drive
is large enough that the effective pinning is 
no longer felt by the rapidly moving particles, and 
implies that if the pinning is absent, the defects will decay in a logarithmic
fashion to a grain boundary dominated state. 
In Fig.~7 we plot $P_{d}(t)$ 
for the same system as in Fig.~6 for $L = 36$ and $f_{p} = 0.0$ where the 
particles were initially placed in random positions. Here 
we observe a logarithmic decay with 
a saturation at long times as the system forms
a grain boundary dominated state. 
This is consistent with the results for high $A^*/f_{p}$ in which
the effective pinning is strongly reduced so the system is closer 
to the clean limit. 
The pinning is necessary for the creation of the smectic 
state and the coarsening of the
domain walls. In the inset we show the same data 
on a log-log scale, indicating that a power
law decay cannot be fit.  
We note that there have been several studies for 
superconducting vortices in two dimensions
interacting with random disorder where a domain 
dominated ground state is observed
\cite{Chandra}. This state only occurs for weak disorder; if 
strong disorder is present, the entire system is disordered. 
In our system the grain boundary
dominated state occurs for weak or no disorder when 
the system is initialized in a random
configuration. In the absence of pinning, the ground state 
would be a completely ordered state; however, the 
grain boundaries we observe do not completely coarsen 
at long times, and the final state of the system
is metastable or glassy.

\begin{figure}
\includegraphics[width=3.5in]{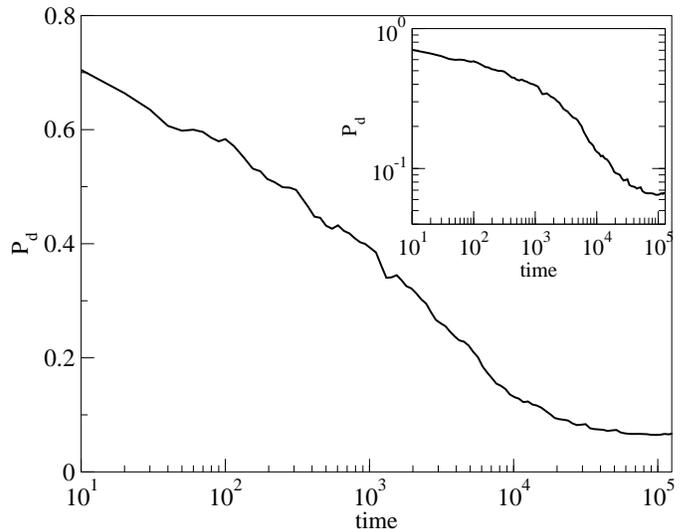}
\caption{
$P_d(t)$ for the same system
as in Fig.~6 with $L = 36$ and no disorder.  
Inset: log-log plot of the
same curve. 
}
\end{figure}

\begin{figure}
\includegraphics[width=3.5in]{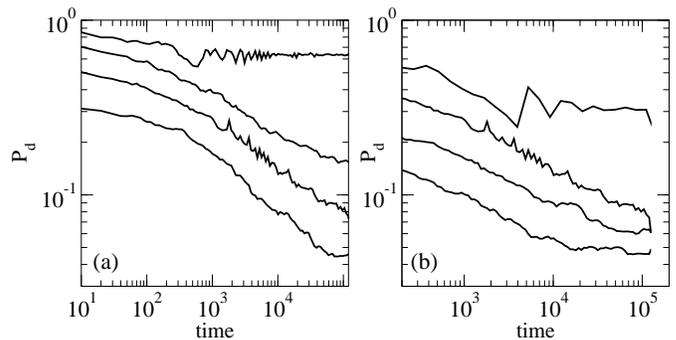}
\caption{
(a) $P_d(t)$ for the same system
as in Fig.~6 with $L = 36$ and $A^*=6$ for varied disorder strength. 
From top to bottom, $A^{*}/f_{p} = 0.7$, 1.25, 2.0, 
and $3.0.$ 
(b) $P_{d}(t)$ for the same system in (a) for varied 
ac frequencies. 
From top to bottom, $\omega/3\times 10^{-4} = 0.1$, 0.25, 1, and 10. 
}
\end{figure}

In Fig.~8(a) we plot the decays for the same system as in Fig.~6 at
$L=36$ with $A^{*}$ held fixed at $A^*=6$ while $f_{p}$ is varied.
This is in contrast to
Fig.~3, where $f_{p}$ was kept fixed at $f_p=1.5$ and $A^{*}$ was varied.  
Here we consider $f_p=8.57$, 4.8, 3.0, and 2.0, corresponding to
$A^*/f_{p} = 0.7$, 1.25, 2.0, and 3.0 in Fig.~8(a). 
For weak disorder, $A^{*}/f_{p} = 3.0$, the decay is
rapid with an exponent $\alpha$ somewhat higher than $0.25$. 
This is consistent with the results
presented in Fig.~2(b) at
$A^{*}/f_{p} = 4.0$ where the effective pinning
is weak but still present and the exponent 
$\alpha=0.33$. 
Higher values of $\alpha$
indicate a more rapid
decay in the defect density. 
When the pinning strength is increased, there is also some creation
of defects caused by the motion over the disordered substrate
which competes with the annihilation of the defects. 
This results in a slower decay of $P_d$, as shown in Fig.~8(a)
for larger pinning forces $A^{*}/f_{p} = 1.25$, 
where $\alpha$ is smaller than $0.25$. 
For the largest pinning force $f_p=8.57$, plotted in
Fig.~8(a) at $A^{*}/f_{p} = 0.7$, 
$P_d$ decays only briefly before 
saturating to a constant value
at which the rate of generation and annihilation
of defects balances. 
These results show that the if the disorder is too strong, 
coarsening of the defects cannot occur.  

In Fig.~8(b) we examine the effects of changing $\omega$ for
the same system as in Fig.~6 with $L = 36$, fixed 
$A^{*}/f_{p} = 2.0$, $A^*=6$, and $f_p=3.0$. 
In Fig.~6 the frequency is $\omega = 3\times 10^{-4}$. 
Fig.~8(b) shows $P_{d}(t)$ 
for $\omega/3\times 10^{-4} = 0.1$, 0.25, 1, and 10, from top to bottom. 
For very low frequencies 
the system remains strongly disordered, as indicated by the
saturation of $P_{d}$ at long times for $\omega/3\times 10^{-4}=0.1$, 
since the particles have time to become partially pinned
during the portion of each drive cycle when the ac amplitude is near zero.   
If the frequency is high enough, as for
$\omega/3\times 10^{-4}\ge 0.25$ in Fig.~8(b),
the system is able to 
coarsen and a power law decay of defect density can occur.
The slope $\alpha$ is somewhat less than $0.25$, 
indicating a relatively slow decay of defect density,
which is consistent with the
fact that some defects are being created even as other defects annihilate.
For the higher values of $\omega$, 
such as for $\omega/3\times 10^{-4}=10$ in Fig.~8(b),
a power law decay of $P_d$ still
occurs; however, there is a cutoff in the coarsening process
at long times. This
may be due to the fact that the effective temperature is dropping
due to the reduced sampling of the pinning landscape, and as a result,
the last of the defects cannot be annihilated away. We note that
at frequencies $\omega/3\times 10^{-4}  = 200$, the effect of the ac drive
is negligible 
and the system decays as if there were no ac drive and only pinning.
In this case the defect density decays 
logarithmically but reaches saturation at much
earlier times than for the case of no pinning illustrated in Fig.~7. 

Our results showing 
logarithmic decay for weak effective disorder and no disorder
are in agreement with the recent experimental results for the 
coarsening of domain
walls in colloidal systems where there was no disorder present. The simulations
by Valenzuela found evidence for a logarithmic decay in the presence of 
quenched disorder \cite{Valenzuela};
however, the scaling range was small and the data was very noisy
so it was not clear if those results indicated a true logarithmic decay.

\section{Summary} 

In summary, we have examined the coarsening of topological
defects in two dimensional systems with quenched disorder when 
an external oscillating drive is suddenly changed from 
a disordered steady state value to
an ordered steady state value. For   
quenches where the final value of the ac drive is close
to the disorder-order crossover, the defect density decays
as a power law in time, $1/t^{\alpha}$, 
with $\alpha = 1/4$ to $1/3$, in agreement with
studies of quenched dynamics in continuum models.
The final state is free of grain boundaries  and the system
aligns in the direction of the ac drive. The remaining
dislocations are also aligned with the direction
of the ac drive and the system forms a smectic state as seen in 
previous simulations by Valenzuela.  If the  final
value of the ac drive is much larger than the
maximum pinning force, the defect density decays in a logarithmic fashion
and the final state is dominated by grain boundaries.
We also find that if the disorder is very weak or absent, the
defect density decreases logarithmically with time.
We
interpret our results according to an effective shaking temperature,
which is low when the ac amplitude is much larger than
the pinning force or when the pinning is absent. In this case,
domain walls of defects are formed which do not completely anneal.  
We specifically
show that these results can be applied to vortices in superconductors
and to colloidal particles.   

We thank M.B. Hastings and W. Zureck for useful discussions. 
This work was supported by the U.S.~Department of Energy under
Contract No.~W-7405-ENG-36.

\end{document}